\documentclass[twocolumn,showpacs,preprintnumbers,amsmath,amssymb,pra]{revtex4}

\usepackage{graphicx}
\usepackage{dcolumn}
\usepackage{bm}

\begin{document}


\title{Light-shift tomography in an optical-dipole trap for neutral atoms}
\author{J.P. Brantut}
\author{J.F. Cl\'ement}
\author{M. Robert de Saint Vincent}
\author{G. Varoquaux}
\author{R.A. Nyman}
\altaffiliation[present address : ]{Center for Cold Matter, Imperial College, London, SW7 2BW, UK}
\author{A. Aspect}
\author{T. Bourdel}
\author{P. Bouyer}
\affiliation{Laboratoire Charles Fabry de l'Institut d'Optique, Univ Paris Sud, CNRS, campus polytechnique RD128 91127 Palaiseau France}

\date{\today}

\begin{abstract}
We report on light-shift tomography of a cloud of $^{87}$Rb atoms in a far-detuned optical-dipole trap at 1565 nm. Our method is based on standard absorption imaging, but takes advantage of the strong light-shift of the excited state of the imaging transition, which is due to a quasi-resonance of the trapping laser with a higher excited level. We use this method to (i) map the equipotentials of a crossed optical-dipole trap, and (ii) study the thermalisation of an atomic cloud by following the evolution of the potential-energy of atoms during the free-evaporation process.
\end{abstract}

\pacs{37.10.Gh,32.60.+i,33.40.+f,32.60.+i}
\maketitle


{\it In-situ} studies of ultracold atomic gases can yield invaluable information. Imaging on the one hand \cite{Andrews07051996,stamperkurn,esteve:130403,ott}, and light-shift spectroscopy on the other hand \cite{PhysRevLett.59.1659,PhysRevLett.91.223001, adams,njp} have independently proved successful in the investigations of cold neutral gases. In this paper, we present a light-shift tomography method combining {\it in-situ} absorption imaging and light-shift spectroscopy to yield an image, and/or the number, of atoms at constant potential-energy in an optical-dipole trap. It takes advantage of the strong light-shifts of the upper $5P_{3/2}$ level of the imaging transition of $^{87}$Rb under the influence of our trapping laser at 1565 nm. This is due to quasi-resonances to higher excited states ($4D_{3/2}$ and $4D_{5/2}$) at 1529 nm. As a consequence, although the light-shift experienced by atoms in the ground state ($5S_{1/2}$), i.e. the trapping potential, is moderate, the shift of the imaging transition can be large compared to its natural linewidth. It thus allows us to perform spectrally-resolved imaging of the atomic cloud.

We use this technique for two different goals: (i) {\it Mapping of the optical potential}. Starting with a cold cloud with a smooth density profile, we suddenly switch on a trapping laser at 1565 nm, and immediately take an absorption image of the atoms in the presence of the trap, before any evolution of the atom density. By repeating this imaging at various probe laser frequencies, we obtain a map of the equal light-shift regions, i.e. we perform tomography of the trap potential \cite{bagnato}. (ii) {\it Measurement of the atom potential-energy distribution}. Counting the number of atoms at a given probe detuning, i.e. at a given potential-energy in the trap, and repeating this measurement at various probe detunings, we directly measure the potential-energy distribution of the cloud. This allows us to study the relaxation of a trapped atomic cloud from an initial out-of-equilibrium situation towards a thermal distribution, by monitoring the evolution of the energy distribution during the free-evaporation process \cite{PhysRevA.64.034703}.

\begin{figure}
\includegraphics[width=0.5\textwidth]{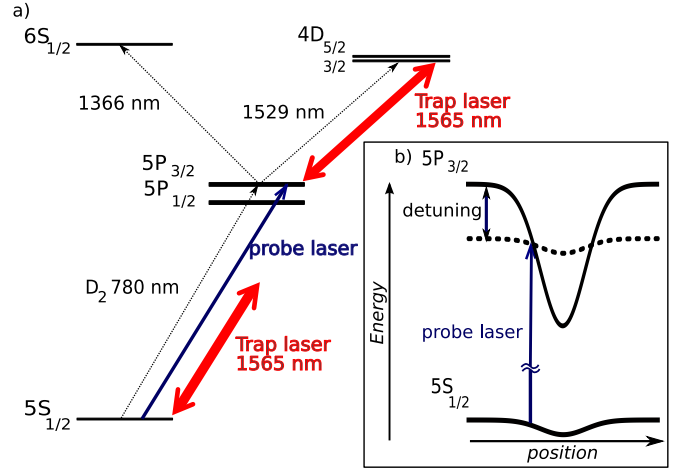}
\caption{\label{fig:Rb} (a): Energy diagram of the lowest energy levels of $^{87}$Rb. The strong transitions at 1529 nm are responsible for the 42.6 enhancement factor in light-shift of the $5P_{3/2}$ energy level with respect to the $5S_{1/2}$ ground state. (b): Light-shifts of the $5S_{1/2}$ and $5P_{3/2}$ modified by a Gaussian focused trap laser at 1565 nm (not on scale). The probe laser interacts with atoms at position which depends an the probe detuning. (color online)}
\end{figure}

As shown in  Figure~\ref{fig:Rb}a, we image $^{87}$Rb atoms using a probe resonant with the $5S_{1/2}$ ($F=2$) to $5P_{3/2}$ ($F'=3$) resonance transition at 780 nm ($D_{2}$), perturbed by the trapping laser at 1565 nm. In order to evaluate the light-shift of the ground and excited states of the $D_{2}$ imaging transition, we have computed the polarisabilities ($\alpha_{g}$ and $\alpha_{e}$, respectively) of these states for the 1565 nm laser excitation. The transitions giving the main contributions are shown in Figure~\ref{fig:Rb}a. The moderate detuning of the trapping laser with respect to the $5P$ to $4D$ transitions, compared to the $5S$ to $5P$, leads to a polarizability of the $5P_{3/2}$ state larger than the one of the ground state by a factor $\frac{\alpha_{e}}{\alpha_{g}}=42.6$.

As the light-shifts of the energy levels are all proportional to the trapping-laser intensity, the light-shift of the ground state, i.e. the potential-energy of an atom in the trap $E_p$ (neglecting gravity), is simply related to the light-shift of the imaging transition $\Delta$ by:

\begin{equation}
\hbar \Delta = (\frac{\alpha_e}{\alpha_g} -1) E_p = 41.6 \, E_p.
\label{eq1}
\end{equation}

As indicated below, we operate in a
situation where the ground state light-shift is of a few MHz only,
whereas the light-shift of the $5P_{3/2}$ level is much larger
than the natural linewidth of the probe transition
$\Gamma=6.1$\,MHz. In facts, when one considers the various magnetic
sublevels of the hyperfine levels $F=2$ and $F'=3$, there is a additional intensity dependant broadening, but it remains
small compared to the light-shift $\Delta$ \cite{hyperfine,clark}. 

As a consequence we can spectrally resolve the potential-energy of the
atoms in the trapping beams. As shown in Figure~\ref{fig:Rb}b, by taking an
{\it in-situ} absorption image with a chosen detuning $\delta$ of
the probe laser, we detect atoms situated at positions $\mathbf{r}$
such that $U({\mathbf{r}}) = \frac{\hbar \delta}{41.6}$, where $U({\mathbf{r}})$ is the
trapping potential ($U = 0$ in free space). We thus directly obtain a map of the trapping potential $U(\mathbf{r})$. The energy resolution is related to the linewidth of the imaging transition, but, as a consequence of Eq.\,\ref{eq1}, enhanced by a factor of 41.6 (corresponding to 0.14\,MHz or 7\,$\mu$K).

The experimental setup is as follows: in a first vacuum chamber,
$^{87}$Rb atoms are collected from an atomic vapour created by a
dispenser in a two-dimensional magneto-optical trap (2D MOT).
Atoms are then transferred into a second chamber where they are
trapped in a 3D MOT. The laser system for 2D and 3D MOT is based
on tapered-amplified extended-cavity diode lasers, as
described in \cite{Papierlaser}. The probe beam is generated by a
dedicated, offset-locked laser, which can be detuned by 350 MHz
from the $F=2$ to $F'=3$ transition, and is linearly polarized.

\begin{figure}
\includegraphics[width = 0.5\textwidth]{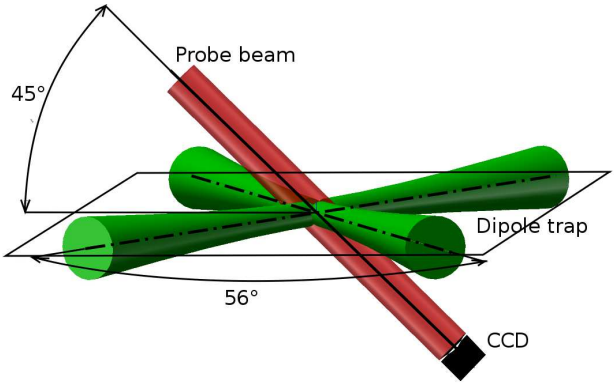}
\caption{\label{fig:beams} Beam configuration of the dipole trap and the probe. (color online)}
\end{figure}

The dipole-trap light is generated by a 50\,W erbium-doped fiber-laser at 1565 nm (IPG ELR-50). This laser is focused to a 50 $\mu \rm m$ waist onto the atomic cloud through a lens that can be moved along the optical axis via a motorized translation stage (Aerotech ANT-50L), in order to be able to vary the beam diameter at the atomic position (much as in ref~\cite{Weiss}). Light is then refocused on atoms by a 1:1 telescope to create a crossed-beam configuration, with a crossing angle of $56^{ \rm o}$. The beam configuration is shown on Fig.\,\ref{fig:beams}. Throughout this paper, the movable lens is placed such that at the dipole trap crossing, the beam diameters are about 200 ${\rm \mu m}$. The optical power is controlled at the output of the laser by an electro-optic modulator and a Glan polarizer. The angle of incidence of the trapping beams with respect to the vacuum-chamber window is $28^{ \rm o}$, leading to astigmatism of the beams. We compensate this effect to first order on the first-pass beam using a tilted, glass plate.

\begin{figure}
\begin{center}
\includegraphics[width=0.155\textwidth]{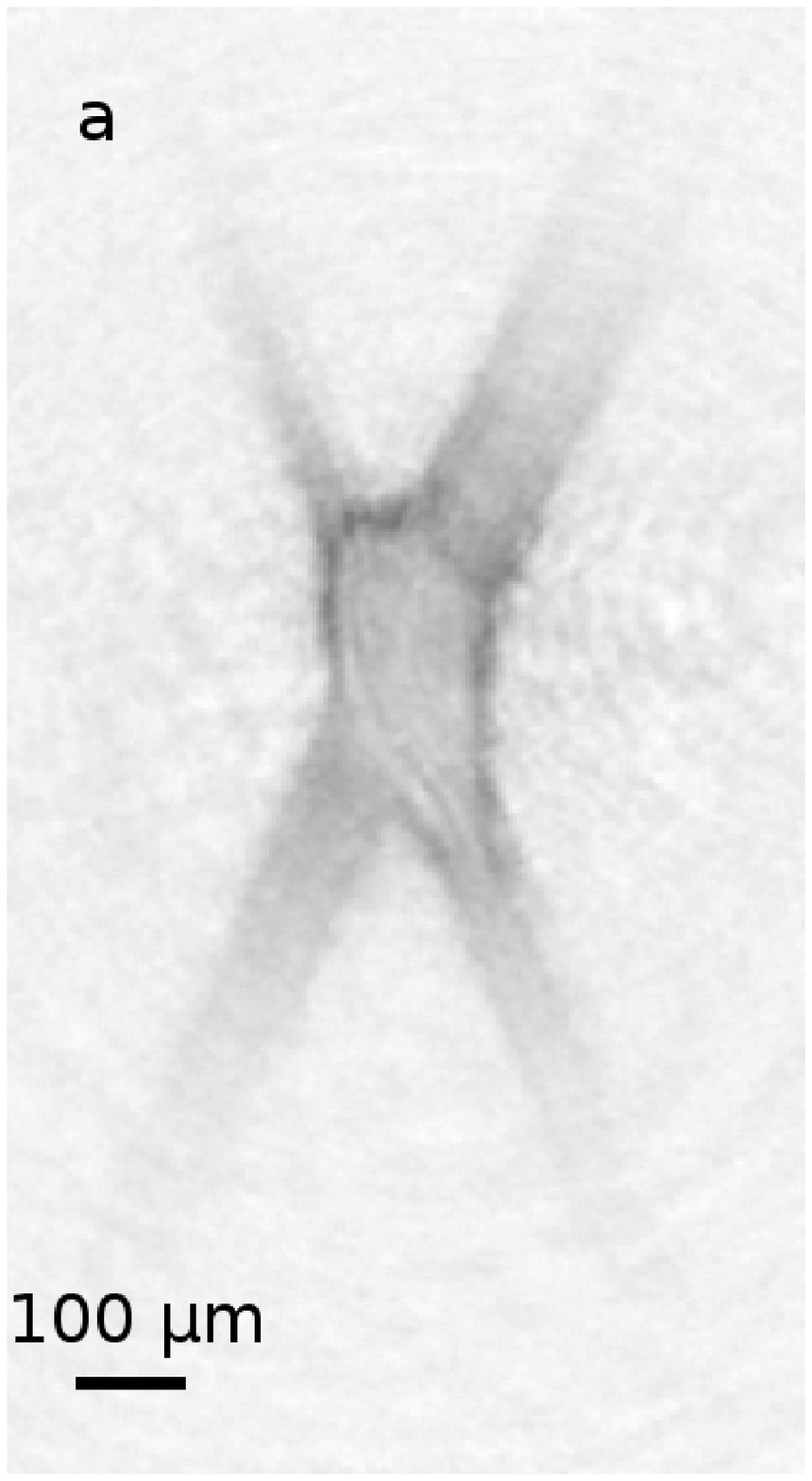} 
\includegraphics[width=0.155\textwidth]{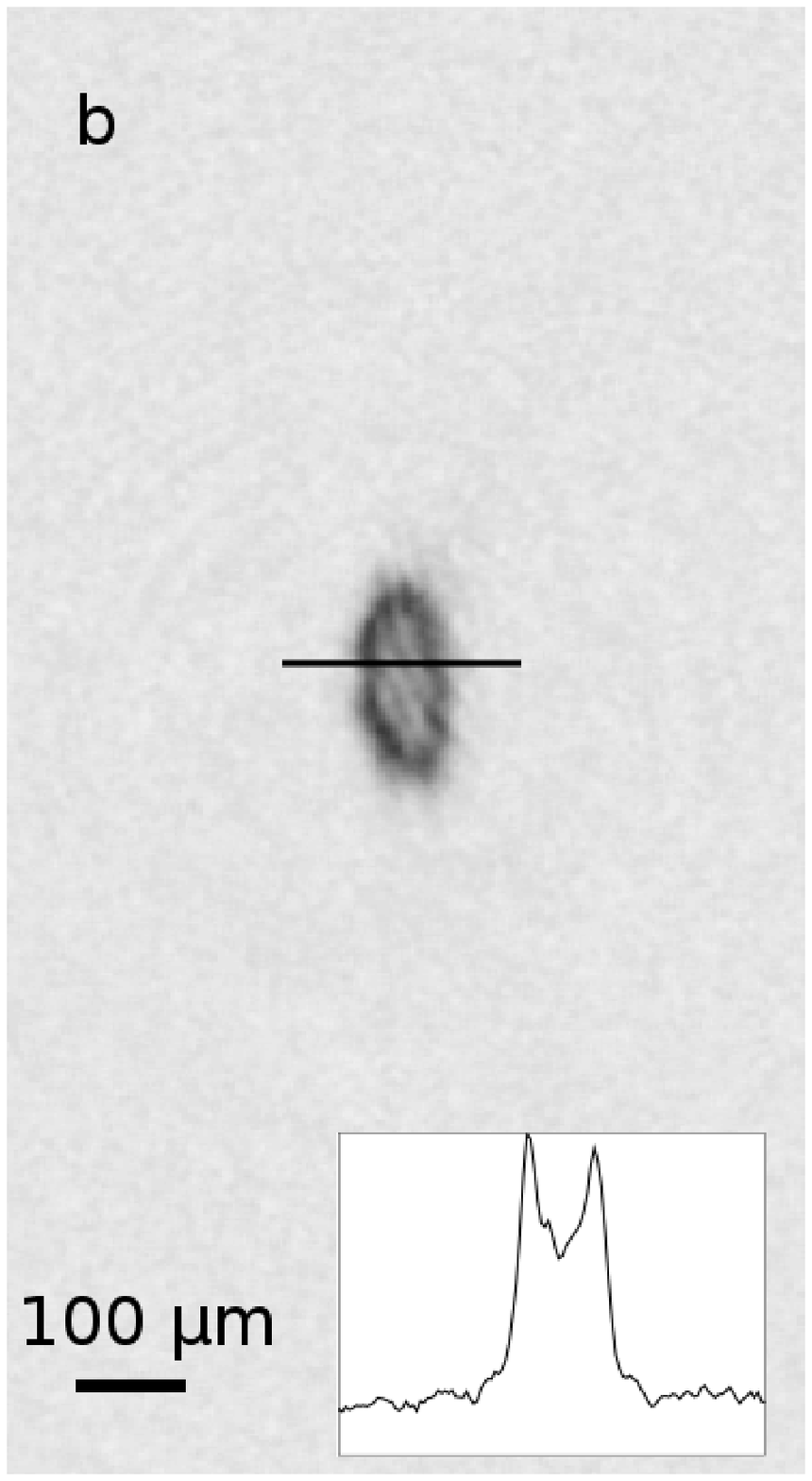} 
\includegraphics[width=0.155\textwidth]{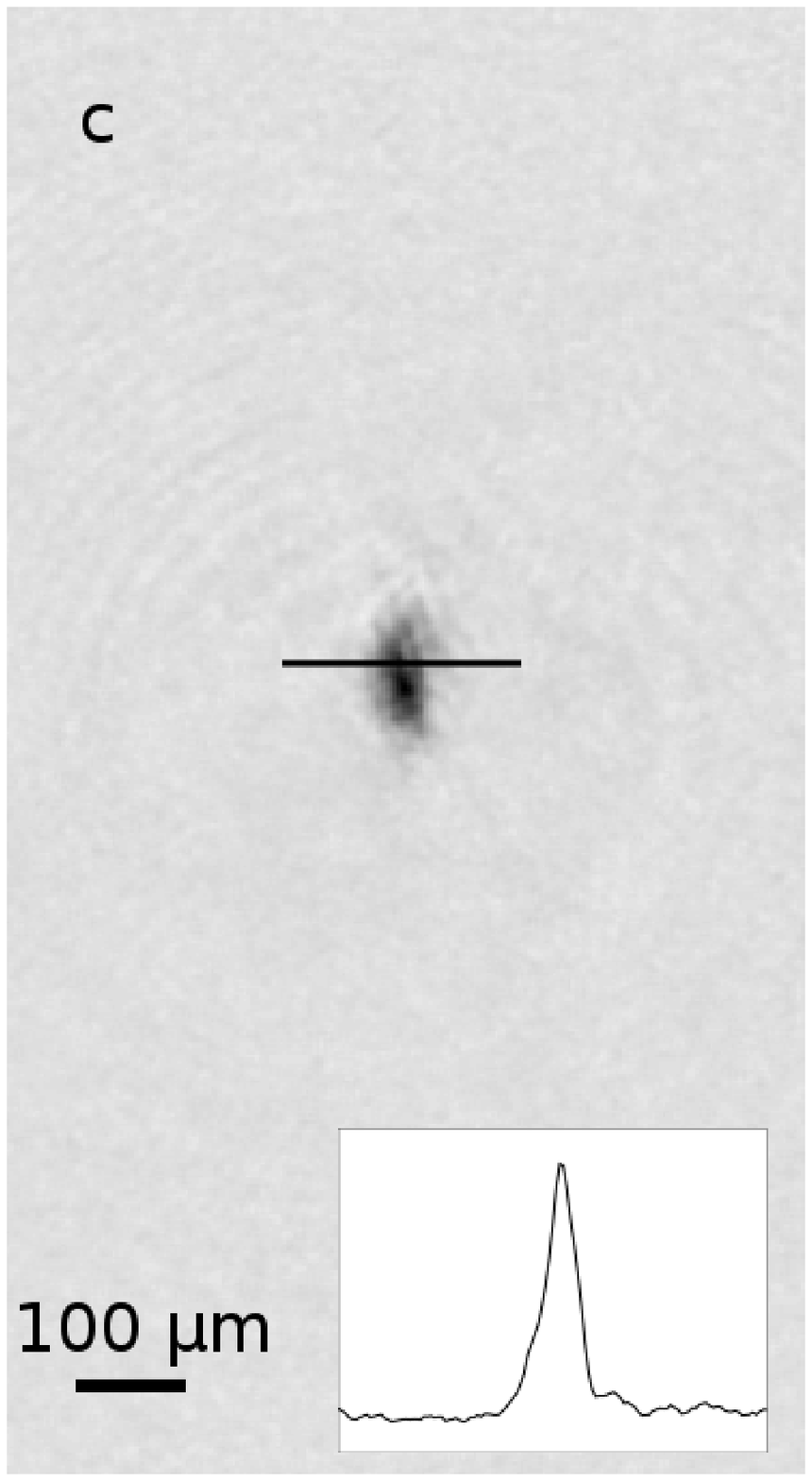} 
\caption{\label{fig:images_map}Absorption images taken immediately after switching on the optical-dipole trap, showing the equipotentials of the trap. By choosing the probe laser detuning with respect to the free-space imaging transition ((a): -40 MHz, (b): -80 MHz, and (c): -100 MHz, corresponding to a light-shift of the ground state of 45 $\mu$K, 91 $\mu$K, and 115 ${\rm \mu K}$ respectively) we observe the distribution of equal laser-intensity regions i.e. equipotentials of the trap. The probe laser propagation axis is inclined by $45^{ \rm o}$ with respect to the crossed dipole trap plane. Images insets (b) and (c) show cuts along the horizontal axis. }
\end{center}
\end{figure}

We use light-shift tomography of the optical-dipole trap to map
the potential landscape, using atoms as local probes of the
potential. About $10^8$ $^{87}$Rb atoms are collected in the 3D
MOT with a magnetic field gradient of $12.5$ $\rm G.cm^{-1}$ and a
detuning of the cooling laser of $2.5 \, \Gamma$. The 3D MOT is
then compressed down to $ 200$ $\rm \mu m$ (rms radius) by
increasing the detuning of the cooling laser up to $10 \, \Gamma$.
The cooling beams are switched off, and the dipole trap laser
is switched on 1 ms later. After another millisecond
an absorption image is taken with a $50$ $\mu s$ pulse at the chosen
probe detuning. A repumping laser resonant with the free space
$F=1$ to $F'=2$ transition is applied together with the imaging
laser. At resonance, atoms scatter about $100$ photons during the
imaging pulse and the atomic density has no time to evolve.

Examples of absorption images are shown in Fig.\,\ref{fig:images_map}. At moderate probe detuning (-40 MHz), we see open equipotentials (Fig.\,\ref{fig:images_map}a), which means that we probe atoms which will be able to escape along the directions of the two beams. In this image, we can detect an asymmetry of the trapping potential between the two arms of the dipole trap, which is related to intensity imbalance and residual astigmatism. The decrease of the contrast at the edges of the image is due to the vanishing atomic density. For larger detunings (-80 MHz), the images (Fig.\,\ref{fig:images_map}b) show closed equipotentials with an elliptic contour. As we go deeper in the trap (-100 MHz), we reach the bottom, where the equipotentials merge into a spot (Fig.\,\ref{fig:images_map}c).

This tomographic method allows us to determine {\it in-situ} the trap characteristics. The potential landscape at the bottom of the trap yields the trap frequencies (110 $\pm$ 10 Hz, 110 $\pm$ 10  Hz, and 150 $\pm$ 10 Hz), and the detuning at which we observe the change from closed to open equipotentials yields the trap depth ($57$ ${\rm \mu K}$ in our experiment). We observe that the shape of the crossed region as probed by tomography is very sensitive to the overlap of the two arms, which provides us with a direct and accurate alignment method \cite{aberrations}.

We evaluate the spatial resolution of the potential mapping as follows: in a place where the trapping potential gradient is $\nabla U$, the resolution associated with the linewidth $\Gamma$ of the probe transition is $d = \frac{\alpha_g}{\alpha_e - \alpha_g} \frac{\hbar \Gamma }{|\nabla U({\bold r})|} \simeq 1/41.6 \frac{\hbar \Gamma }{|\nabla U({\bold r})|}$. In our experimental conditions, at the places of steepest gradients, $d$ =10 ${\rm \mu m}$, whereas the resolution of our imaging setup is 7 ${\rm \mu m}$. The widths of the peaks in the insets of figures \ref{fig:images_map}b and c are explained by the convolution of the two effects. Actually, $d$ could be made smaller than the resolution of the imaging system, either by increasing the light-shift gradients or by using a narrower transition \cite{thomasth,thomasexp}.


We now turn to the measure of the potential-energy distribution of
the atomic cloud. For each value of the probe detuning, i.e. of the potential-energy (Eq.\,\ref{eq1}), we count
the number of atoms by integration of the optical density in the corresponding
{\it in-situ} absorption image (any spatial information is
lost). Repeating this measurement at various values of the detuning, we obtain the atomic potential-energy
distribution convoluted with our energy resolution lineshape. As
in our case the measured atom number varies smoothly at the scale
of the resolution (7\,$\mu$K), it is simply proportional to the
atomic potential-energy distribution.

Using this technique, we follow the evolution of an optically-trapped cloud under the effect of free evaporation. Experimentally, we load atoms in a compressed MOT, we then turn off the cooling laser, and load the atoms in the dipole trap \cite{loading_JF}. We then wait for a given time, and image {\it in-situ} the trapped cloud at various detunings. 

\begin{figure}
\includegraphics[width = 0.5\textwidth]{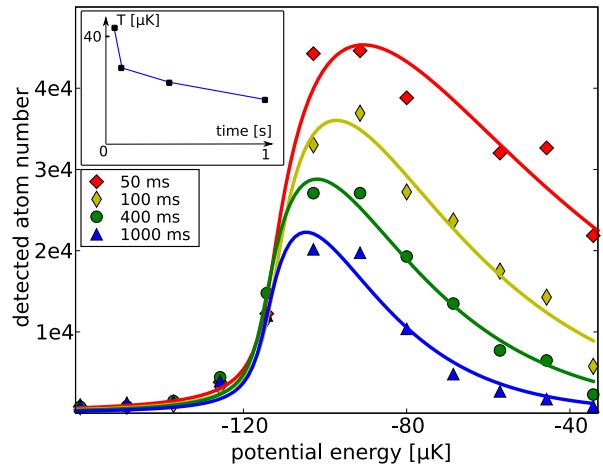}
\caption{\label{fig:thermalisation} Thermalisation of a trapped atomic cloud. The number of detected atoms in the trap is measured as a function of potential-energy, after different evolution times. The potential-energy is taken to be zero in free-space. Solid curves are fits for a thermal gas taking into account the finite linewidth of the imaging transition (using equation~\ref{fiteq}). Fitted temperatures are 41, 27, 22 and 16 $\mu K$, consistent with time-of-flight measurements. Inset shows the evolution of the fitted temperature of the cloud as a function of time. (color online)}
\end{figure}

Figure~\ref{fig:thermalisation} shows the observed atomic
potential-energy distribution after various evolution times,
averaged over 4 images per data point. At short evolution times
($<$50\,ms), the measurement is blurred by the untrapped atoms
falling from the magneto-optical trap. After 50\,ms, the atom
distribution extends far into the anharmonic region of the trap,
in particular into the two arms where atoms can escape. The fact
that we can observe atoms with a potential-energy larger than the
trap depth is a signature of the non-ergodicity of the
free-evaporation process: atoms have enough energy to escape but
temporarily remain in the trapping region. Not all the available
regions of phase-space are explored in the time scale of the
experiment \cite{evap}. The bottom of the trap is at -115\,$\mu
\rm K$. Atoms detected below this value reflect the finite linewidth of the
imaging transition \cite{hyperfine}.

As the thermalisation proceeds, the number of atoms located in regions of high potential-energy diminishes faster than the number of atoms at the bottom of the trap. This process results in a lower mean potential-energy of the remaining atoms. To characterize our results better, we fit our data with the theoretically predicted atom number $N(\delta)$ at a given probe detuning, in the case of $N_{\rm tot}$ classical atoms at thermal equilibrium in a harmonic trap, taking into account the finite linewidth of the imaging transition:
\begin{equation}
\label{fiteq}
N(\delta) =  \frac{4N_{\rm tot}}{\sqrt{\pi}} \int_{0}^{\infty}
\frac{u^2e^{-u^2} du}{1+4 (\delta + v -tu^2)^2}
\end{equation}
where $t = \frac{k_B T}{\hbar
\Gamma} (\frac{\alpha_e} { \alpha_g} -1 )$, $v=\frac{U(r = 0)}{\hbar
\Gamma}  \frac{\alpha_e} { \alpha_g}$ and $\delta = \frac{\omega - \omega_0}{\Gamma}$ ($\omega_0$ is the unshifted probe transition frequency) are normalized temperature, minimal potential-energy and detunings respectively. The fit is unexpectedly good given the fact that the trap is harmonic only close to the bottom of the trap. The fitted temperatures are, to within 10\,$\%$ uncertainty consistent with complementary measurements of the velocity distribution obtained by the time-of-flight method.

Inset in Figure~\ref{fig:thermalisation} shows the evolution of
the temperature with the free evaporation time. We first observe a
rapid decrease of the temperature in 100\,ms, which we attribute
to a simple loss of atoms with an energy larger than the trap depth. It takes a few trap
periods for the energetic atoms to find their path to escape
along one of the two trapping beams. At longer time, we observe a
slower decrease of the temperature. It is consistent with
collision-induced evaporation as we estimate the initial collision
rate in our trap to be of the order of 3\,s$^{-1}$. The lifetime
in this experiment is limited to 2\,s by background gas collisions
and we nevertheless measure an increase in phase-space density by
a factor of 5 between the loaded cloud at 50\,ms and the thermal,
trapped cloud at equilibrium after 1\,s.

In conclusion we have presented a method to directly map the potential created on $^{87}$Rb by an erbium doped fiber-laser at 1565 nm and to directly measure the atomic potential-energy distribution. The energy resolution ($7$ ${\rm \mu K}$) could be improved using a trapping laser closer to the 1529 nm transition and an appropriate probe polarisation \cite{hyperfine}. This method is not specific to $^{87}$Rb, and could be generalized to other alkali gases which have strong transitions from the $nP_{3/2}$ state to higher excited states in the mid-infrared (e.g. Cs at 1469 nm, K at 1252 nm) \cite{nist}.

The potential-energy resolution of this technique is independent of the spatial resolution of the imaging setup. For a rapidly spatially-varying light field, this technique could achieve sub-wavelength resolution. For example, in a three-dimensional, cubic, optical lattice, the detuned probe laser would be resonant with atoms situated in spherical shells centered on each lattice site, the radius of which being related to the probe-laser detuning. Light-shift tomography, allowing {\it in-situ} local probing, is likely to be useful in the the context of both quantum gas studies and quantum information processing using cold atoms.

We acknowledge technical assistance from F. Moron and A. Villing. This research was supported by CNRS, CNES as part of the ICE project, the project "blanc" M\'elaBoF\'erIA from ANR, IFRAF ; by the STREP program FINAQS of the European Union and by the MAP program SAI of the European Space Agency (ESA).


\begin{thebibliography}{20}
\expandafter\ifx\csname natexlab\endcsname\relax\def\natexlab#1{#1}\fi
\expandafter\ifx\csname bibnamefont\endcsname\relax
  \def\bibnamefont#1{#1}\fi
\expandafter\ifx\csname bibfnamefont\endcsname\relax
  \def\bibfnamefont#1{#1}\fi
\expandafter\ifx\csname citenamefont\endcsname\relax
  \def\citenamefont#1{#1}\fi
\expandafter\ifx\csname url\endcsname\relax
  \def\url#1{\texttt{#1}}\fi
\expandafter\ifx\csname urlprefix\endcsname\relax\def\urlprefix{URL }\fi
\providecommand{\bibinfo}[2]{#2}
\providecommand{\eprint}[2][]{\url{#2}}

\bibitem[{\citenamefont{Andrews et~al.}(1996)\citenamefont{Andrews, Mewes, van
  Druten, Durfee, Kurn, and Ketterle}}]{Andrews07051996}
\bibinfo{author}{\bibfnamefont{M.~R.} \bibnamefont{Andrews}},
  \bibinfo{author}{\bibfnamefont{M.-O.} \bibnamefont{Mewes}},
  \bibinfo{author}{\bibfnamefont{N.~J.} \bibnamefont{van Druten}},
  \bibinfo{author}{\bibfnamefont{D.~S.} \bibnamefont{Durfee}},
  \bibinfo{author}{\bibfnamefont{D.~M.} \bibnamefont{Kurn}}, \bibnamefont{and}
  \bibinfo{author}{\bibfnamefont{W.}~\bibnamefont{Ketterle}},
  \bibinfo{journal}{Science} \textbf{\bibinfo{volume}{273}},
  \bibinfo{pages}{84} (\bibinfo{year}{1996}).

\bibitem[{\citenamefont{Higbie et~al.}(2005)\citenamefont{Higbie, Sadler,
  Inouye, Chikkatur, Leslie, Moore, Savalli, and Stamper-Kurn}}]{stamperkurn}
\bibinfo{author}{\bibfnamefont{J.~M.} \bibnamefont{Higbie}},
  \bibinfo{author}{\bibfnamefont{L.~E.} \bibnamefont{Sadler}},
  \bibinfo{author}{\bibfnamefont{S.}~\bibnamefont{Inouye}},
  \bibinfo{author}{\bibfnamefont{A.~P.} \bibnamefont{Chikkatur}},
  \bibinfo{author}{\bibfnamefont{S.~R.} \bibnamefont{Leslie}},
  \bibinfo{author}{\bibfnamefont{K.~L.} \bibnamefont{Moore}},
  \bibinfo{author}{\bibfnamefont{V.}~\bibnamefont{Savalli}}, \bibnamefont{and}
  \bibinfo{author}{\bibfnamefont{D.~M.} \bibnamefont{Stamper-Kurn}},
  \bibinfo{journal}{Phys. Rev. Lett.} \textbf{\bibinfo{volume}{95}},
  \bibinfo{eid}{050401} (\bibinfo{year}{2005}).

\bibitem[{\citenamefont{Esteve et~al.}(2006)\citenamefont{Esteve, Trebbia,
  Schumm, Aspect, Westbrook, and Bouchoule}}]{esteve:130403}
\bibinfo{author}{\bibfnamefont{J.}~\bibnamefont{Esteve}},
  \bibinfo{author}{\bibfnamefont{J.-B.} \bibnamefont{Trebbia}},
  \bibinfo{author}{\bibfnamefont{T.}~\bibnamefont{Schumm}},
  \bibinfo{author}{\bibfnamefont{A.}~\bibnamefont{Aspect}},
  \bibinfo{author}{\bibfnamefont{C.~I.} \bibnamefont{Westbrook}},
  \bibnamefont{and}
  \bibinfo{author}{\bibfnamefont{I.}~\bibnamefont{Bouchoule}},
  \bibinfo{journal}{Phys. Rev. Lett.} \textbf{\bibinfo{volume}{96}},
  \bibinfo{eid}{130403} (\bibinfo{year}{2006}).

\bibitem[{\citenamefont{Gericke et~al.}(2008)\citenamefont{Gericke, Wurtz,
  Reitz, Langen, and Ott}}]{ott}
\bibinfo{author}{\bibfnamefont{T.}~\bibnamefont{Gericke}},
  \bibinfo{author}{\bibfnamefont{P.}~\bibnamefont{Wurtz}},
  \bibinfo{author}{\bibfnamefont{D.}~\bibnamefont{Reitz}},
  \bibinfo{author}{\bibfnamefont{T.}~\bibnamefont{Langen}}, \bibnamefont{and}
  \bibinfo{author}{\bibfnamefont{H.}~\bibnamefont{Ott}} (\bibinfo{year}{2008}),
  \bibinfo{note}{arXiv.org:0804.4788}.

\bibitem[{\citenamefont{Salomon et~al.}(1987)\citenamefont{Salomon, Dalibard,
  Aspect, Metcalf, and Cohen-Tannoudji}}]{PhysRevLett.59.1659}
\bibinfo{author}{\bibfnamefont{C.}~\bibnamefont{Salomon}},
  \bibinfo{author}{\bibfnamefont{J.}~\bibnamefont{Dalibard}},
  \bibinfo{author}{\bibfnamefont{A.}~\bibnamefont{Aspect}},
  \bibinfo{author}{\bibfnamefont{H.}~\bibnamefont{Metcalf}}, \bibnamefont{and}
  \bibinfo{author}{\bibfnamefont{C.}~\bibnamefont{Cohen-Tannoudji}},
  \bibinfo{journal}{Phys. Rev. Lett.} \textbf{\bibinfo{volume}{59}},
  \bibinfo{pages}{1659} (\bibinfo{year}{1987}).

\bibitem[{\citenamefont{Takamoto and Katori}(2003)}]{PhysRevLett.91.223001}
\bibinfo{author}{\bibfnamefont{M.}~\bibnamefont{Takamoto}} \bibnamefont{and}
  \bibinfo{author}{\bibfnamefont{H.}~\bibnamefont{Katori}},
  \bibinfo{journal}{Phys. Rev. Lett.} \textbf{\bibinfo{volume}{91}},
  \bibinfo{pages}{223001} (\bibinfo{year}{2003}).

\bibitem[{\citenamefont{Griffin et~al.}(2006)\citenamefont{Griffin, Weatherill,
  MacLeod, Potvliege, and Adams}}]{adams}
\bibinfo{author}{\bibfnamefont{P.~F.} \bibnamefont{Griffin}},
  \bibinfo{author}{\bibfnamefont{K.~J.} \bibnamefont{Weatherill}},
  \bibinfo{author}{\bibfnamefont{S.~G.} \bibnamefont{MacLeod}},
  \bibinfo{author}{\bibfnamefont{R.~M.} \bibnamefont{Potvliege}},
  \bibnamefont{and} \bibinfo{author}{\bibfnamefont{C.~S.} \bibnamefont{Adams}},
  \bibinfo{journal}{New J. Phys.} \textbf{\bibinfo{volume}{8}},
  \bibinfo{pages}{11} (\bibinfo{year}{2006}).

\bibitem[{\citenamefont{Cl\'{e}ment et~al.}(2006)\citenamefont{Cl\'{e}ment,
  Var\'{o}n, Retter, Sanchez-Palencia, Aspect, and Bouyer}}]{njp}
\bibinfo{author}{\bibfnamefont{D.}~\bibnamefont{Cl\'{e}ment}},
  \bibinfo{author}{\bibfnamefont{A.~F.} \bibnamefont{Var\'{o}n}},
  \bibinfo{author}{\bibfnamefont{J.~A.} \bibnamefont{Retter}},
  \bibinfo{author}{\bibfnamefont{L.}~\bibnamefont{Sanchez-Palencia}},
  \bibinfo{author}{\bibfnamefont{A.}~\bibnamefont{Aspect}}, \bibnamefont{and}
  \bibinfo{author}{\bibfnamefont{P.}~\bibnamefont{Bouyer}},
  \bibinfo{journal}{New J. Phys.} \textbf{\bibinfo{volume}{8}},
  \bibinfo{pages}{165} (\bibinfo{year}{2006}).

\bibitem[{\citenamefont{Courteille et~al.}(2001)\citenamefont{Courteille,
  Muniz, Magalhaes, Kaiser, Marcassa, and Bagnato}}]{bagnato}
\bibinfo{author}{\bibfnamefont{P.}~\bibnamefont{Courteille}},
  \bibinfo{author}{\bibfnamefont{S.}~\bibnamefont{Muniz}},
  \bibinfo{author}{\bibfnamefont{K.}~\bibnamefont{Magalhaes}},
  \bibinfo{author}{\bibfnamefont{R.}~\bibnamefont{Kaiser}},
  \bibinfo{author}{\bibfnamefont{L.}~\bibnamefont{Marcassa}}, \bibnamefont{and}
  \bibinfo{author}{\bibfnamefont{V.}~\bibnamefont{Bagnato}},
  \bibinfo{journal}{Eur. Phys. J. D} \textbf{\bibinfo{volume}{15}},
  \bibinfo{pages}{173} (\bibinfo{year}{2001}).

\bibitem[{\citenamefont{Browaeys et~al.}(2001)\citenamefont{Browaeys, Robert,
  Sirjean, Poupard, Nowak, Boiron, Westbrook, and Aspect}}]{PhysRevA.64.034703}
\bibinfo{author}{\bibfnamefont{A.}~\bibnamefont{Browaeys}},
  \bibinfo{author}{\bibfnamefont{A.}~\bibnamefont{Robert}},
  \bibinfo{author}{\bibfnamefont{O.}~\bibnamefont{Sirjean}},
  \bibinfo{author}{\bibfnamefont{J.}~\bibnamefont{Poupard}},
  \bibinfo{author}{\bibfnamefont{S.}~\bibnamefont{Nowak}},
  \bibinfo{author}{\bibfnamefont{D.}~\bibnamefont{Boiron}},
  \bibinfo{author}{\bibfnamefont{C.~I.} \bibnamefont{Westbrook}},
  \bibnamefont{and} \bibinfo{author}{\bibfnamefont{A.}~\bibnamefont{Aspect}},
  \bibinfo{journal}{Phys. Rev. A} \textbf{\bibinfo{volume}{64}},
  \bibinfo{pages}{034703} (\bibinfo{year}{2001}).

\bibitem[{hyp()}]{hyperfine}
\bibinfo{note}{The tensor polarizability leads to a difference of the
  polarizability of the various magnetic sublevels of the $5P_{3/2}$ ($F'=3$)
  state of about 20\,$\%$ \cite{clark}. Due to the orthogonal polarizations of
  the two arms of the crossed dipole trap, the polarization of the trapping
  laser is ill defined in the trap, and we assume a uniform distribution of the
  atoms over the various magnetic sublevels of the ground level $5S_{1/2}$
  ($F=2$). The broadening of the probe transition is then evaluated by taking
  equal polarization components for the probe beams. This results into an
  intensity-dependant broadening that can be as large as 12\,$\%$ of the shift
  of the transition, which adds to the natural linewidth (6.1\,MHz).}

\bibitem[{\citenamefont{Arora et~al.}(2007)\citenamefont{Arora, Safronova, and
  Clark}}]{clark}
\bibinfo{author}{\bibfnamefont{B.}~\bibnamefont{Arora}},
  \bibinfo{author}{\bibfnamefont{M.~S.} \bibnamefont{Safronova}},
  \bibnamefont{and} \bibinfo{author}{\bibfnamefont{C.~W.} \bibnamefont{Clark}},
  \bibinfo{journal}{Phys. Rev. A} \textbf{\bibinfo{volume}{76}},
  \bibinfo{eid}{052509} (\bibinfo{year}{2007}).

\bibitem[{\citenamefont{Nyman et~al.}(2006)\citenamefont{Nyman, Varoquaux,
  Villier, Sacchet, Moron, Le~Coq, Aspect, and Bouyer}}]{Papierlaser}
\bibinfo{author}{\bibfnamefont{R.}~\bibnamefont{Nyman}},
  \bibinfo{author}{\bibfnamefont{G.}~\bibnamefont{Varoquaux}},
  \bibinfo{author}{\bibfnamefont{B.}~\bibnamefont{Villier}},
  \bibinfo{author}{\bibfnamefont{D.}~\bibnamefont{Sacchet}},
  \bibinfo{author}{\bibfnamefont{F.}~\bibnamefont{Moron}},
  \bibinfo{author}{\bibfnamefont{Y.}~\bibnamefont{Le~Coq}},
  \bibinfo{author}{\bibfnamefont{A.}~\bibnamefont{Aspect}}, \bibnamefont{and}
  \bibinfo{author}{\bibfnamefont{P.}~\bibnamefont{Bouyer}},
  \bibinfo{journal}{Rev. Sci. Instrum.} \textbf{\bibinfo{volume}{77}},
  \bibinfo{pages}{033105} (\bibinfo{year}{2006}).

\bibitem[{\citenamefont{Kinoshita et~al.}(2005)\citenamefont{Kinoshita, Wenger,
  and Weiss}}]{Weiss}
\bibinfo{author}{\bibfnamefont{T.}~\bibnamefont{Kinoshita}},
  \bibinfo{author}{\bibfnamefont{T.}~\bibnamefont{Wenger}}, \bibnamefont{and}
  \bibinfo{author}{\bibfnamefont{D.~S.} \bibnamefont{Weiss}},
  \bibinfo{journal}{Phys. Rev. A} \textbf{\bibinfo{volume}{71}},
  \bibinfo{eid}{011602(R)} (\bibinfo{year}{2005}).

\bibitem[{abe()}]{aberrations}
\bibinfo{note}{Using the potential mapping, we can evaluate {\it in-situ} the
  aberrations created on the trap beam by the vaccum chamber windows, by
  measuring the real beam waist, and extracting the beam quality factor
  M${^2}$.}

\bibitem[{\citenamefont{Thomas}(1990)}]{thomasth}
\bibinfo{author}{\bibfnamefont{J.~E.} \bibnamefont{Thomas}},
  \bibinfo{journal}{Phys. Rev. A} \textbf{\bibinfo{volume}{42}},
  \bibinfo{pages}{5652} (\bibinfo{year}{1990}).

\bibitem[{\citenamefont{Gardner et~al.}(1993)\citenamefont{Gardner, Marable,
  Welch, and Thomas}}]{thomasexp}
\bibinfo{author}{\bibfnamefont{J.~R.} \bibnamefont{Gardner}},
  \bibinfo{author}{\bibfnamefont{M.~L.} \bibnamefont{Marable}},
  \bibinfo{author}{\bibfnamefont{G.~R.} \bibnamefont{Welch}}, \bibnamefont{and}
  \bibinfo{author}{\bibfnamefont{J.~E.} \bibnamefont{Thomas}},
  \bibinfo{journal}{Phys. Rev. Lett.} \textbf{\bibinfo{volume}{70}},
  \bibinfo{pages}{3404} (\bibinfo{year}{1993}).

\bibitem[{loa()}]{loading_JF}
\bibinfo{note}{The strong red-shift of the cycling transition raises specific
  problems in the loading process from a MOT or an optical molasses. Here we
  address this problem by rapidly alterning the optical dipole trap and the
  cooling lasers (10\,kHz). Efficient loading of an optical-dipole trap at
  1565\,nm will be described in a further publication.}

\bibitem[{\citenamefont{Ketterle and Van~Druten}(1996)}]{evap}
\bibinfo{author}{\bibfnamefont{W.}~\bibnamefont{Ketterle}} \bibnamefont{and}
  \bibinfo{author}{\bibfnamefont{N.}~\bibnamefont{Van~Druten}},
  \bibinfo{journal}{Adv. At. Mol. Opt. Phys.} \textbf{\bibinfo{volume}{37}}
  (\bibinfo{year}{1996}).

\bibitem[{nis()}]{nist}
\bibinfo{note}{Ralchenko, Yu., Kramida, A.E., Reader, J. and NIST ASD Team
  (2008). NIST Atomic Spectra Database (version 3.1.4). National Institute of
  Standards and Technology, Gaithersburg, MD.}

\end{thebibliography}
\end{document}